# Step bunching during Si(001) homoepitaxy caused by the surface diffusion anisotropy

J. Myslivecek[1], C. Schelling[2], F. Schäffler[2], G. Springholz[2], P. Šmilauer[3], J. Krug[4], B. Voigtländer[1]

[1]Institut für Schichten und Grenzflächen, Forschungszentrum Jülich, D-52425 Jülich, Germany
[2]Institut für Halbleiterphysik, Johannes Kepler Universität, A-4040 Linz, Austria
[3]Czech Academy of Sciences, Cukrovarnická 10, 162 53 Praha 6, Czech Republic
[4]Fachbereich Physik, Universität Essen, D-45117 Essen, Germany

**Abstract:**
Scanning tunneling microscopy experiments show that the unstable growth morphology observed during molecular beam homoepitaxy on slightly vicinal Si(001) surfaces consists of straight step bunches. The instability occurs under step-flow growth conditions and vanishes both during low-temperature island growth and at high temperatures. An instability with the same characteristics is observed in a 2D Kinetic Monte Carlo model of growth with incorporated Si(001)-like diffusion anisotropy. This provides strong evidence that the diffusion anisotropy destabilizes growth on Si(001) and similar surfaces towards step bunching. This new instability mechanism is operational without any additional step edge barriers.

Morphological instabilities of thin films during growth and annealing are the basic mechanisms for the preparation of self-assembled quantum dots and wires, but they are detrimental to the interface quality of semiconductor heterostructures. Hence, a detailed understanding of the underlying mechanisms is essential for either application. Up to now, most investigations in this field assumed the dominance of strain-driven phenomena, but it becomes more and more obvious that kinetic step bunching is a widespread instability during homo- and heteroepitaxy [1,2]. In this paper we argue that the kinetic step bunching on vicinal Si(001) during molecular beam homoepitaxy (MBE) [3] is an inherent property of the 2×1 reconstructed Si(001) surface. In particular, kinetic Monte Carlo (KMC) simulations confirm that this step bunching instability appears in a growth model with Si(001)-like diffusion anisotropy without any additional step edge barriers. This new instability mechanism is expected to be relevant for all semiconductor surfaces of similar symmetry.

In our experiment, 1000 Å thick silicon epilayers were deposited by MBE at 0.16 Å/s on Si(001) substrates miscut by 0.66° along [110]. After deposition, the samples were quenched to room temperature and transferred in situ to the STM. We tested that the thermal cleaning by direct current produces flat surfaces with equally distributed $S_A$ and $S_B$ steps [4].

Surface morphologies after growth at different temperatures are shown in Fig. 1. At 670 K and below, we observe a weak undulation of the surface (< 5 Å, Fig. 1a). Two dimensional (2D) islands grow on the terraces (Fig. 1d). With increasing temperature, a narrow temperature interval is reached (around 760 K for the miscut and the deposition rate employed), where the instability becomes most pronounced (Fig. 1b). A quasi-periodic array of step bunches develops with corrugation amplitudes around 15 Å. The bunches consist of trains of closely arranged $S_B$-$S_A$ step pairs [5, 6]. They are separated by single height terraces much wider than the original terrace width. On these terraces some island nucleation still takes place (Fig. 1e). At even higher temperatures, the step train becomes more regular (Fig. 1c). The corrugation decreases to <7 Å, and island nucleation is no longer observed (Fig. 1f). The STM data provide clear evidence that the ripple morphology is due to a step bunching instability. Ripple formation by electromigration [7] can be ruled out, because the same morphology was also found on radiation heated samples [3]. Since the instability vanishes after high temperature annealing, it is of kinetic origin.

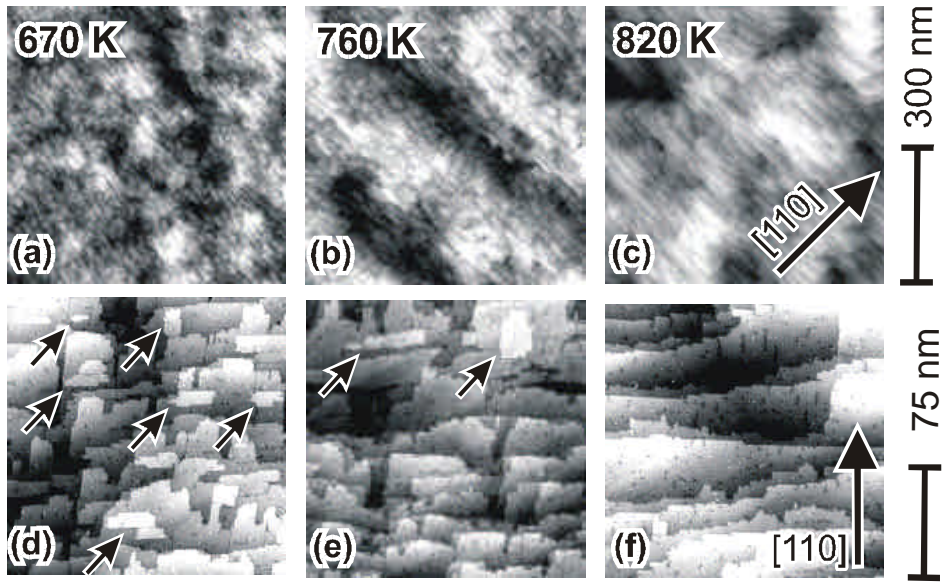

**Figure 1:** Images of 1000 Å Si layers grown at different temperatures on Si(001) miscut 0.66° toward [110]. The height scale is (a) 5 Å, (b) 15 Å, (c) 7 Å. Step bunching is observed around 760 K (b). On the small scale images, island growth is observed at 670 K (d), residual island growth on wide terraces at 760 K (e) and pure step-flow growth at 820 K (f). Arrows in (d,e) indicate 2D island positions.

Growth instabilities under ideal MBE conditions can be discussed in terms of a miscut dependent nonequilibrium surface mass current *j(m)* [8,9]. The miscut direction *m* is locally perpendicular to the step edges and its absolute value is defined as the ratio of step height and average terrace width. The projection of *j* onto the miscut direction *j(m)* (*m*>0 for upward surface staircase going from left to right) identifies three basic linear instabilities: step bunching [9,10] and step meandering [11] during step-flow growth, and growth of mounds [9,12] during island growth. Step meandering is observed for an uphill current (*j*>0) [11], while step bunching or growth of mounds occurs when $\partial j/\partial m \equiv j'(m) > 0$ [9,12]. This classification enables surface stability analyses for any surface potential via extraction of the surface current in a KMC simulation [8,12].

To define a model surface diffusion potential we consider the 2×1 reconstructed Si(001) surface miscut along [110] that consists of alternating A and B terraces separated by single-height steps [5]. Dimer rows of 2×1 reconstruction on A terraces are oriented perpendicular, on B terraces parallel to *m*. The respective downhill step edges are labeled $S_A$ and $S_B$ (Fig. 2a). The surface diffusion is highly anisotropic with diffusion along the dimer rows much faster than across the dimer rows (an order of magnitude at 1000 K) [13,14]. In addition, $S_B$ step edges represent more effective traps for adparticles than $S_A$ step edges [4-6,15,16]. Despite the fact that diffusion of material on the Si surface and its attachment/detachment at step edges is complicated and presumably involves many particle processes [17], considerable insight has been gained by minimal growth models [18,19] where diffusing material units just obey basic diffusion and sticking rules of the Si(001)2×1 surface.

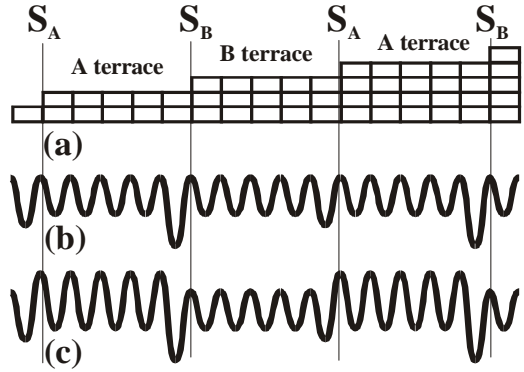

**Figure 2:** Cross section of a Si(001) surface miscut in [110] direction and related cross sections of 2D model surface diffusion potentials. B terraces are wider than A terraces as usual on Si(001) surface during step-flow growth. (b) Si(001)-like potential with isotropic diffusion [18,19] (c) Si(001)-like potential with anisotropic diffusion.

Many characteristic features of Si(001) homoepitaxy were explained by models assuming only different trap potentials at $S_A$ and $S_B$ steps without invoking diffusion anisotropy [18,19], Fig. 2b. These models correctly describe the initially faster propagation of the $S_B$ steps that leads to a relative B terrace coverage $q_B > 50$ %. After deposition of a few monolayers (ML) a stationary state is reached with equal propagation rates of $S_A$ and $S_B$ steps. Then, $q_B$ becomes constant at >50 % (Fig. 2a), and close pairs of $S_B$ and $S_A$ steps, and to some extent double-height $D_B$ step segments, form [5,6]. However, no net surface current is observed in the stationary state, and thus growth is inherently stable in such models.

A surface current during deposition on a vicinal surface requires an asymmetry in the miscut direction of the surface potential around step edge traps. An additional Ehrlich-Schwoebel barrier (ES), which is found on many metal surfaces, would cause such an asymmetry, but since it hampers step feeding from above it opposes step bunching [10]. In this picture a barrier to attachment from the lower terrace (inverse ES [iES]) is required for step bunching [20]. However, there is little evidence for the presence of either type of barriers at the SA and SB step edges [16].

On the other hand, an obvious asymmetry of the step edge traps results by just adding the diffusion anisotropy to the simple model potential of Fig. 2b, as shown in Fig. 2c. To find out whether this modification causes a destabilizing current during deposition, we extracted the surface currents from 2D KMC simulations with this model potential. We used a solid-on-solid growth model on a cubic lattice derived from Ref. 19. Initially, we tested a model with both the anisotropy in diffusion and the different stickiness of $S_A$ and $S_B$ step edges being present. However, it turned out that the diffusion anisotropy alone yields qualitatively the same results regarding growth stability. This is because diffusion anisotropy contributes to a higher kink density at $S_B$ steps [21], and thus causes a difference in the effective stickiness of the two step types.

In the following we will therefore concentrate on the role of diffusion anisotropy. An Arrhenius behavior $\mathbf{n} = \mathbf{n}_0 \exp(-E_A/kT)$ was assumed for the hopping rate of material units. The diffusion barrier has the form $E_A=E_S+nE_N$ $(+E_{ani})$ where $E_S$ is a surface contribution, $E_N$ is a lateral neighbor contribution, $n$ is the number of nearest lateral neighbors before a hop and $E_{ani}$ is added for hops that are perpendicular to the direction of the dimer rows (the direction changes by 90° on neighboring terraces). When a $D_B$ ($D_A$) double step segment is formed in the simulation by an $S_B$ step catching up with an $S_A$ step, the step edge potential of the segment becomes symmetric. The model was tested to obey detailed balance.

Plots of $j(m)$ in the miscut direction for different temperatures are displayed in Fig. 3. Growth simulations took place on initially ideally miscut surfaces with equally distributed $S_A$ and $S_B$ steps. The surface current was averaged between 2 and 7 ML of deposition. The first 2 ML were left out because during this period $S_B$-$S_A$ step pairs form and the surface morphology is not stationary. At low temperatures and miscuts growth proceeds by nucleation of islands, the current is downstairs [$j(m)<0$] and growth is stable [$j'(m)<0$]. As temperature and/or miscut increases, the growth mode changes to step flow by $S_B$-$S_A$ step-pair propagation. Under these conditions, we still find $j(m)<0$ but growth is unstable toward step bunching since $j'(m)>0$. The stationary values of $\mathbf{q}_B$ are higher than 50 % and they decrease with increasing temperature. At the highest temperatures and/or miscuts, both the current and its derivative approach zero and growth is stable. The growth mode remains step flow, and $\mathbf{q}_B$ fluctuates around 50 %.

In agreement with the experiments, we find for any particular miscut three growth modes: Stable island growth at low temperatures, unstable step-pair flow at intermediate temperatures, and stable step flow at high temperatures. Fig. 4 shows cross-sections of the surface morphology obtained from simulations at different temperatures.

At intermediate temperatures/miscuts separately determined surface currents on A and B terraces show that the main contribution to the current comes from B terraces, where the direction of fast diffusion is parallel to the miscut. The current is formed mainly by deposited particles that detach from the straight segments of $S_A$ steps with a strong preference to go downstairs in the presence of diffusion anisotropy (cf. Fig. 2c). This detachment process resembles the reflection of adatoms approaching a step-pair from the lower terrace, and thus effectively behaves like a destabilizing iES. The contribution of A terraces to the current in miscut direction is negligible, because the fast diffusion on A terraces is perpendicular to the miscut.

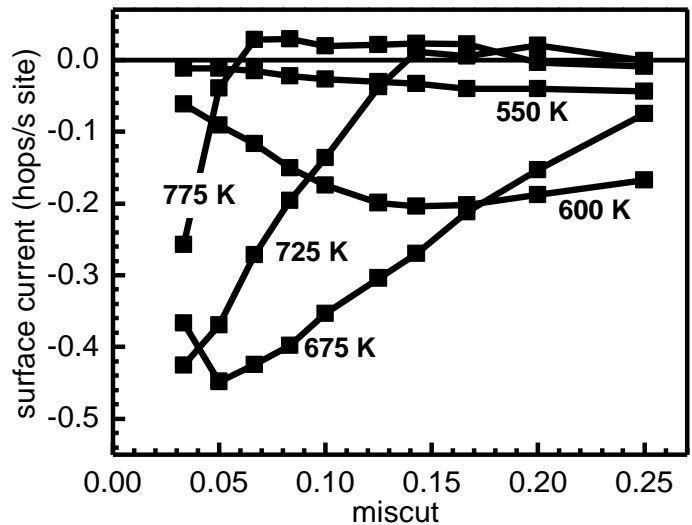

**Figure 3:** Surface current during deposition as a function of miscut extracted from KMC simulations with the potential from Fig. 2c. A positive slope of the curves indicates step bunching. For any particular miscut a temperature interval exists, where the instability is observed.

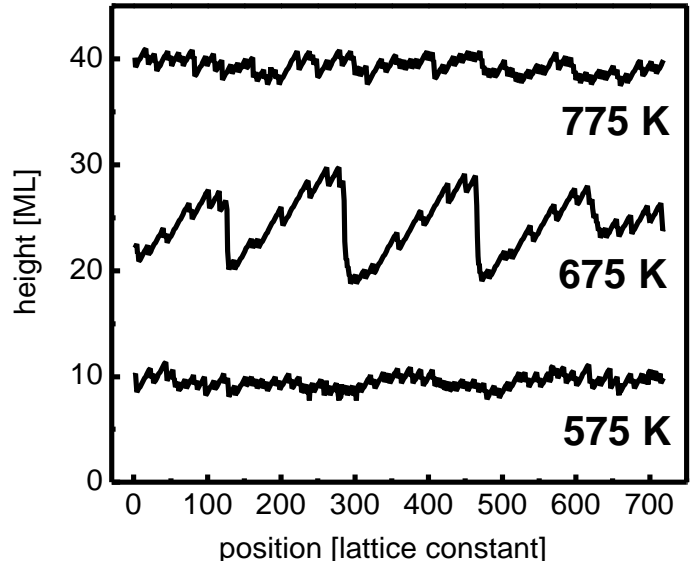

**Figure 4:** Simulated surface cross sections in the miscut direction after 500 ML deposition at miscut=0.125 show the existence of the step bunching instability in this temperature interval. (b), (c) Top view of the unstable surface morphology: step bunching initially occurs locally (b), step bunches merge at later times (c).

At low temperatures/miscuts, island formation on the terraces sets in which changes the sign of $j'(m)$ and stabilizes growth. This is an analogy of the stable/unstable growth transition of a vicinal surface with normal ES barriers [12]. At high temperatures/miscuts, the surface approaches thermal equilibrium. Without deposition attachment/detachment of particles at step edges keeps the steps apart and generates no surface current. This will be also the case during growth, as long as the detachment rate of adatoms from step edges is much faster than the attachment rate of deposited particles. Otherwise, the growth will be kinetically determined by the surface diffusion potential implying the formation of step pairs and nonzero surface current in our case.

To summarize, our KMC simulations show, that a simple 2D growth model with Si(001)-like diffusion anisotropy causes a step bunching instability of the same character as observed in our STM experiment. We could show that in the narrow temperature window, where the instability is observed, the interplay between diffusion anisotropy and step-edge detachment behaves like an effective (and temperature-dependent) iES, even in the absence of any physical step edge barriers. This new generic mechanism is expected to apply to all surfaces of similar symmetry.

We gratefully acknowledge discussions with M. Kotrla. J. M. would like to thank the Alexander von Humboldt Foundation for the support of his stay in Jülich. The work was supported by FWF Vienna (grant P12143PHY), GACR Czech Republic, (grant 202/01/0928), and Volkswagenstiftung (grant I/72872). A more detailed account of the investigations of the kinetic step bunching instability on Si(001) can be found in Ref. 22.